\documentclass[11pt]{article}
\title{Competitive Boolean Function Evaluation:\\ Beyond Monotonicity, and the Symmetric Case}
\author{{\sc Ferdinando Cicalese}\\
\texttt{cicalese@dia.unisa.it}\\University of Salerno,\\ Italy \and
{\sc Travis Gagie}\\\texttt{travis.gagie@gmail.com}\\University of Chile,\\ Chile \and {\sc Eduardo Laber}\\\texttt{laber@inf.puc-rio.br}\\PUC-Rio, \\ Brazil  \and
{\sc Martin Milani\v{c}}\thanks{Corresponding author.}\\\texttt{martin.milanic@upr.si}\\University of Primorska, \\ Slovenia}
\usepackage{amsthm, amsmath, amssymb}

\begin{document}
\maketitle
\newcommand{\spc}{{\vskip 0.15cm}{\noindent}}
\newtheorem{claim}{Claim}
\newtheorem{theorem}{Theorem}
\newtheorem{cor}{Corollary}
\newtheorem{proposition}{Proposition}
\newtheorem{conjecture}{Conjecture}
\newtheorem{lem}{Lemma}
\newtheorem{lemma}{Lemma}
\newtheorem{obs}{Observation}
\newtheorem{definition}{Definition}
\newtheorem{problem}{Problem}
\newcommand{\prf}{\noindent {\bf Proof.~}}
\newcommand{\comment}[1]{}
\newcommand{\todo}[1]{\marginpar{#1}}
\newcommand{\A}{\mathbb{A}}

\newcommand{\remove}[1]{}
\newcommand{\done}{\rule{2mm}{2mm} \vskip \belowdisplayskip}
\newcommand{\commento}[1]{\marginpar{\tiny \flushleft{#1}}}

\def\y{{\bf y}}
\def\p{{\bf p}}
\def\q{{\bf q}}
\def\z{{\bf z}}
\def\u{{\bf u}}

\def\w{{\bf w}}
\def\s{{\bf s}}
\def\a{{\bf a}}
\def\b{{\bf b}}
\newcommand{\B}{\mathbb{B}}
\newcommand{\G}{\mathbb{G}}
\newcommand{\I}{\mathcal{I}}
\newcommand{\F}{\mathcal{F}}
\newcommand{\M}{\mathcal{M}}

\begin{abstract}
We study the extremal competitive ratio of Boolean function evaluation.
We provide the first non-trivial lower and upper bounds for classes of Boolean
functions which are not included in the class of  monotone Boolean functions.
For the particular case of symmetric functions our bounds are matching and
we exactly characterize the best possible competitiveness achievable by a
deterministic algorithm. Our upper bound is obtained by a simple
polynomial time algorithm.
\end{abstract}

\section{Introduction}
A Boolean function $f$ has to be evaluated for a fixed but unknown
choice of the values of the variables.
Each variable $x$ of $f$ has an associated cost $c(x)$,
which has to be paid to read the value of $x$.
The problem is to design algorithms that evaluate the
function querying the values of the variables
sequentially while trying to minimize the total cost incurred.
The evaluation of the performance of an algorithm is done  by
employing competitive analysis, i.e., by considering the ratio between what the
algorithm pays and the cost of the cheapest set of variables needed to certify the
value taken by  the function on the given assignment.

The problem is related to the well studied area of
decision tree complexity of Boolean functions \cite{HeiWig91,JayKumSiv03,HeiNewWig93,karchmer,FOCS::SaksW1986,Snir:1985:LBP}.
Also,  algorithms for efficient evaluation of Boolean functions  play an important role in
several areas, e.g.,  electrical engineering (the analysis and design
of switching networks \cite{Sw_Net}), artificial intelligence (neural networks \cite{Neu_Net}), medicine
(testing patients for a disease), automatic diagnosis \cite{Bor99}, reliability theory \cite{Reliab}, game theory (weighted majority games
\cite{Games}), distributed computing systems (mutual exclusion mechanism \cite{Davidson,Molina:85}, synchronizing
processes \cite{henderson:88}), just to mention a few. We refer to the excellent monograph \cite{Crama-Hammer} for many more examples.

When the function to evaluate is restricted to be monotone, it is known that for any deterministic algorithm there exists a
choice of the costs such that the  best possible competitiveness achievable by the algorithm cannot be smaller than
the maximum size of a certificate (minterm or maxterm) of the function to evaluate.
For a given $f,$ the size of such a  largest certificate is usually referred to as $\textit{PROOF}(f).$
The existence of a (possibly exponential time) algorithm with competitiveness $\textit{PROOF}(f)$ was shown in  \cite{CicLab_icalp}.
Also there exists a polynomial time algorithm\footnote{Here we mean that the algorithm uses a polynomial number of
calls to an oracle which, given an assignment,  provides the value of the function in that  assignment.}
 whose competitiveness is at most  $2 \times \textit{PROOF}(f)$ \cite{CicLab}.
$\textit{PROOF}(f)$-competitive polynomial time algorithms are known for important subclasses of the monotone Boolean functions
\cite{CharikarEtAl02a,CicLab_soda,Cic-Mil}.  Whether $\textit{PROOF}(f)$-competitive polynomial algorithms exist for any monotone $f$ is
 still open. However, it has been observed by the authors that the linear programming based approach of \cite{CicLab_icalp} is
 $\textit{PROOF}(f)$-competitive for any (not necessarily monotone) Boolean function $f.$ This raises an immediate question regarding
 lower bounds for the case of arbitrary Boolean functions. To the best of our knowledge there are no results for the general case of Boolean functions nor
 upper bounds better than $\textit{PROOF}(f).$
In \cite{CicLab_icalp} it is observed that for  arbitrary Boolean functions, $\textit{PROOF}(f)$ can be larger than the  extremal competitiveness.
However, no concrete results are known so far.

In this paper we start the study of arbitrary (non-monotone) Boolean functions in the context of computing with priced information.
We provide the first non-trivial upper and lower bounds on the
competitiveness achievable for some subclasses of non-monotone functions.
 In particular, we give a complete characterization for the case of symmetric functions.
 The (extremal) competitiveness\footnote{A formal definition will be given in Section 2.} only depends on the structural properties of the function to evaluate, therefore it can be considered as a possible measure of the complexity of the function.  According to I. Wegener~\cite{Wegener-sym}, ``\dots the class of symmetric functions includes many fundamental functions, among them all types of counting functions
 [\dots] hence it is a fundamental problem in computer science to determine the complexity of symmetric functions with respect to different models
 of computation''.  Some classical results can be found in \cite{MP,W1984,Bar}.
 Symmetric functions are also considered a fundamental class in the theory of
uniform distribution learning   \cite{Blum,Blum2}.
The model we use here can also be thought of as a special type of learning (see, e.g., \cite{Kaplan2005}).

\medskip
{\bf The structure of the paper.}
In Section 2, we formally define the problem and the quantity involved in the analysis of the evaluation algorithms.
In Section 3 we consider the class of quadratic Boolean functions, i.e., Boolean functions with a DNF only including terms of cardinality at most two. In \cite{CicLab} a
$\textit{PROOF}(f)$-competitive algorithm was given for any monotone function in this class. It is not hard to see that
the same algorithm achieves competitiveness  $\textit{PROOF}(f)$ also in the case of non-monotone functions.
In the light of the above observations, we study the problem of lower bounds  for such class of functions.
We show that for the class of quadratic Boolean functions, denoted by ${\cal Q}$, it holds that $2 \leq \sup_{f \in {\cal Q}} \textit{PROOF}(f) / \gamma(f) \leq 3,$ where $\gamma(f)$ denotes the best
extremal competitiveness achievable for $f.$
In other words the algorithm of \cite{CicLab} provides an approximation of a factor between two and three of the optimal extremal competitiveness.

In Section 4 we provide an optimal algorithm for the class of symmetric Boolean functions. Besides exactly evaluating the
extremal competitiveness for the class, we provide a polynomial algorithm which  achieves
the best possible competitiveness with respect to any fixed cost assignment.

Finally, in Section 5, we discuss some possible future directions of research. We
provide some technical results allowing us to determine the extremal competitiveness for Boolean functions having a DNF in which some variables only appear in negated or non-negated form. We believe that this result, which is based on the linear programming approach of
\cite{CicLab_icalp}, could give some insight on a possible approach of reducing Boolean function evaluation to monotone Boolean function evaluation, via some factorization.

\section{Preliminaries}

In this section, we formally define the function evaluation problem and the
quantity involved in the analysis of the evaluation algorithms. A function
$f(x_1, \dots, x_n)$ has to be evaluated for a fixed but unknown choice of
the values for the set of variables $V=\{x_1,x_2,\ldots,x_n\}$.
Each variable  $x_i$ has an associated non-negative cost $c({x_i})$
which is the cost incurred to probe $x_i$, i.e.,
to read its value. Given a set $U \subseteq V$, we define the
cost $c(U)$ of $U$ as the sum of the costs of the variables in $U$,
i.e., $c(U) = \sum_{x \in U} c(x).$
The goal is to {\em adaptively} identify and probe a
minimum cost set of variables $U \subseteq V$ whose values uniquely
determine the value of $f,$ regardless of
the value of the variables not probed.

An {\it assignment } $\sigma$ for a function $f$ is a choice of a
value for each of its variables.
We shall denote by  $x_i(\sigma)$ the value
assigned to $x_i$ in the assignment $\sigma.$
We use $f(\sigma)$ to denote the value of $f$ w.r.t.~$\sigma$,
i.e., $f(\sigma) = f(x_1(\sigma), \dots, x_n(\sigma)).$
Given a subset $U \subseteq V,$ we use $\sigma_{U}$ to denote the
restriction of $\sigma$ to the variables in $U.$
Given an assignment $\sigma,$ we say that $U$ is a {\em proof} of
$f$ for the assignment $\sigma$ if
the value of  $f$ is determined by the partial assignment
$\sigma_{U}$ and this is not true for any subset of $U.$
More generally, we say that  a set of variables $U$ is
a {\em proof } of $f$ if there exists an assignment for which $U$ is a proof of $f$.
By $\textit{PROOF}(f)$ we denote the maximum size of a proof of $f$.

An evaluation algorithm $\A$ for $f$ is a decision tree, that is,
a rule to adaptively read the variables in $V$ until the set of
variables read so far includes a proof for the value of $f$. The
cost of algorithm $\A$ for an assignment $\sigma$ is the total
cost incurred by $\A$ to evaluate $f$ under the assignment $\sigma$.
Given a cost function $c(\cdot),$ we let  $c^f_{\A}(\sigma)$ denote
 the cost of the algorithm $\A$
for an assignment $\sigma$ and $c^f(\sigma)$ the cost of the
cheapest proof for $f$ under the assignment $\sigma.$ We say that
$\A$ is $\rho$-competitive if $  c^f_{\A}(\sigma) \leq \rho
c^f(\sigma),$ for every possible assignment $\sigma$. We use
$\gamma_c^{\A}(f)$ to denote the competitive ratio of $\A$ with respect
to the cost $c(\cdot)$, that is,
the infimum over all values of $\rho$ for which $\A$ is $\rho$-competitive. The best
possible competitive ratio for any deterministic algorithm, then, is
\[ \gamma_c^{f} = \inf_{\A } \gamma_c^{\A}(f),\] where the infimum is
taken over all possible deterministic algorithms $\A$.

With the aim of evaluating the dependence of the competitive ratio
on the structure of $f$, the  extremal competitive ratio
$\gamma^{\A}(f)$ of an algorithm $\A$ is defined as \[ \gamma^{\A}(f) = \sup_{c}
\gamma_c^{\A}(f)\,,\] where the supremum is taken over all non-negative cost
functions $c:V\to \mathbb{R}^+$. The best possible extremal  competitive ratio for
any deterministic algorithm, then, is \[ \gamma(f) = \inf_{\A }
\gamma^{\A}(f).\] This last measure is meant to capture the
structural complexity of $f$ independent of a particular cost
assignment and algorithm.

We denote the set $\{0,1\}$ by $\B$.

\section{Quadratic Boolean functions}\label{sect:quadr}

\vspace{-0.2truecm}

For a Boolean function $f$ on a set of
$n$ variables
$V=\{x_1,x_2,\ldots,x_n\}$, a literal either
refers to a variable
$x_i$ or to its negation $\overline{x_i}$.
In order to define the
value of a literal, we
proceed as follows.
Given an assignment
$\sigma$ and a variable $x$, we define the
value of the negation of
$x$ by $\overline{x}(\sigma)
=1- x(\sigma).$
Accordingly, by fixing the value of a literal to a value $v$,
we mean to fix the value of the corresponding variable $x$ to $1-v$,
if the literal coincides with $\overline{x}$ and to fix the value of $x$ to $v$, otherwise.

A minterm (maxterm) for $f$ is a minimal
set of literals such that
if we set the values of all its literals
to 1 (0) then $f$ evaluates
to 1 (0)  regardless of the values assigned
to the other variables.

In this section we  restrict our analysis to the case of {\em quadratic} Boolean functions. These are the Boolean functions every minterm of which has size at most two. This class coincides with the class of Boolean function admitting a DNF
in which all the terms are of size at most~2 (see, e.g.,~\cite{Crama-Hammer}).
We shall denote the class of quadratic Boolean functions by ${\cal Q}$.

We use $k(f)$ and $l(f)$ to denote the size of the largest minterm and the largest maxterm of $f,$ respectively. Therefore, for quadratic Boolean functions, we have $\max\{k(f), l(f)\} = \textit{PROOF}(f).$



It is known
that $\gamma(f) = \max\{k(f),l(f)\},$
holds for {\em every} {\em
monotone} Boolean function $f.$
It is also known that $\gamma(f) \leq  \max\{k(f),l(f)\},$
holds for the whole class of Boolean functions \cite{CicLab_icalp,CicLab}.
\remove{
Moreover, there is a
polynomial time algorithm that achieves
competitive ratio $
2\max\{k(f),l(f)\}$ in the general
case and competitive ratio
$\max\{k(f), l(f)\}$ when $f$ is
a monotone function that belongs to
${\cal Q}$ \cite{CicLab}. A careful examination of
this algorithm shows that it
extends to every function in ${\cal Q}$ in the sense that its
analysis ensures
 $\gamma(f) \leq \max\{k(f),l(f)\}$,
 for every $ f \in {\cal Q}$.
 }
A natural question arising from the above results is
about the lower bound on the
competitiveness of any algorithm
that evaluates functions in $\cal Q$.
In this section we shall provide a partial answer to that question by
concentrating  on the following
issue:

{\em What is the maximum $K$ such that, for every $f \in
{\cal Q}$, $\gamma(f) \geq
K \max\{k(f),l(f)\}$?}

The next two results are to the
 effect that $1/3 \leq K \leq 1/2$ .

\begin{theorem}
For each $f \in {\cal Q}$, we
have $ \gamma(f) \geq l(f) / 3$.
\end{theorem}
\begin{proof}
Let $\cal F$ be the family of
 the minterms of $f$. Given a set of
literals $U$, we use $var(U)$ to
 denote the variables of $U$. Let
$C$ be a largest  maxterm for $f$ and
let $ L $ be the subset of
literals of $C$ defined by
$L = \left\{ \ell \in C \mid \mbox{either }
\{ \ell \} \in {\cal F}
\mbox{ or } \{ \ell, \overline{m} \} \in {\cal F},
\mbox{for some literal } m \in C \right\}$.

Let $\sigma'$ be an assignment for
the variables of $V \setminus
var(C)$. We say that a literal
$\ell \in C$  survives to  $\sigma'$ if
$f(\sigma)=1$ for every assignment
 $\sigma$ such that
 $\sigma_{V \setminus var(C)} = \sigma'$ and
$\ell(\sigma)=1$. Now, let $\sigma^*$
be the assignment of the
variables of $V \setminus var(C)$
that maximizes the number of
literals of $C \setminus L$ that
survive. Let $L^*$ be the subset of literals in $C
\setminus L$ which survive to~$\sigma^*.$

{\bf Claim.} $|L^*| \geq |C-L|/2$.


To see this, we first notice that each literal $\ell \in C\setminus L$ appears in
a minterm together with some literal $m$ such that $var(\{m\}) \not \subseteq var(C).$
Indeed: by the definition of $L$, for each literal $\ell \in C \setminus L,$ every minterm which contains $\ell$ can only contain a literal from $V \setminus C$ or a literal $\ell'$ such that $\ell' \in C.$ However, it cannot happen that  $\ell$ appears only in minterms together with  some other literal from $C,$ for otherwise  $C$ would not be minimal.

Now, let $m$ be a literal such that  $var(\{m\}) \not \subseteq var(C)$ and $\{\ell, m\}$ is a minterm. Clearly, for every
assignment $\sigma'$ for $V \setminus var(C)$ such that $m(\sigma') = 1$ the literal $\ell$
survives to $\sigma'.$

We can now construct the assignment $\sigma^*$ for $V \setminus var(C)$ as follows.
For each variable $x \in  V \setminus var(C),$ let
$C_1(x) = \{ \ell \in C \setminus L \mid \{x,\ell\} \mbox{ is a minterm}\}$ and
$C_0(x) = \{ \ell \in C \setminus L \mid \{\overline{x},\ell\} \mbox{ is a minterm}\}.$ In words,
$C_0(x)$ ($C_1(x)$) is  the set of literals of $C \setminus L$ which appear in a minterm together with $x$ ($\overline{x}).$
We set $x(\sigma^*) = 1$ if and only if $|C_1(x)| > |C_0(x)|.$
From the above observations, we have that $C\setminus L = \bigcup_{x \in V \setminus var(C)} C_1(x) \cup C_0(x).$
Let $C_{\max}(x)$ be the larger set between $C_1(x)$ and $C_0(x).$
We also have $$|L^*| = \left|\bigcup_{x \in V \setminus var(C)} C_{\max}(x)\right|  \geq
\frac{1}{2} \left| \bigcup_{x \in V \setminus var(C)} C_1(x) \cup C_0(x) \right| = \frac{1}{2} |C \setminus L|,$$
which concludes the proof of the claim.

Let $c_1(\cdot)$ be the cost map
 defined by $c_1(x)=1$ for each
variable $x \in var(L \cup L^*)$
and $c_1(x)=0$, otherwise. Fix an
algorithm $\A$ that evaluates $f$.
Let $\sigma_{\A}$ be the
assignment defined by:
$x(\sigma_{{\A}}) = x(\sigma^*),$ for each $x
\in V \setminus var(C)$ and
 $\ell(\sigma_{{\A}}) = 0$ for every
literal $\ell \in C,$\ but for
the last one that ${\A}$ reads in  $L^*
\cup L$. It is not hard to verify that  $f(\sigma_{\A})=1.$
 Indeed, if the last variable read is from $L^*,$
 then it corresponds to a literal which survives to $\sigma^*,$ whence
 by setting to $1$ we have $f = 1,$   by the definition of surviving literal.
On the other hand, if the last variable read by $\A$ is from $L,$ then
it corresponds to a literal $\ell$  which is alone in some minterm, or to a literal that is
in some minterm together with the negation of another literal $m$ from $L.$
Since $\sigma_{\A}$ is such that $\ell(\sigma_{\A}) =1 $ and $m(\sigma_{\A}) =0 $
we have again $f = 1.$
From the above two cases it is also easy to see that while the
algorithm incurs a cost of
 $|L| +|L^*|$ to evaluate $f,$
the cost
of the cheapest proof is at most $2$.
Thus, we have $\gamma_{c_1}^f
\geq (|L| +|L^*|)/2$.

Now, consider the cost map
 $c_2(\cdot)$ defined by:
  $c_2(x)=1$ for
each variable $x$ corresponding
to a literal in $ L^*$ and
$c_2(x)=0$, otherwise.
Proceeding as before, for each algorithm $\A$
that evaluates $f$ let $\sigma_{\A}$
 be the assignment for the
variables of $f$ such that
$x(\sigma_{\A}) = x(\sigma^*)$ for each
$x \in V \setminus var(C)$ and
  $\ell(\sigma_{\A}) = 0$ for each
literal $\ell$ in $C$, but for the last
one of $L^*$ read by $\A$.
Therefore, such an instance forces the
algorithm to incur a cost equal to
 $|L^*|$ whilst the cost of the
  cheapest proof is $1$.
Thus, $\gamma_{c_2}^f \geq |L^*|$.

Putting together the above
results, we have $\gamma(f) \geq
\max\{\gamma_{c_1}^f,
\gamma_{c_2}^f\} \geq \max \{|L^*|, (|L|
+|L^*|)/2 \} \geq |C|/3$,
where the least inequality follows from
$|L^*| \geq |C -L|/2$.
\end{proof}

In order to complete our partial
answer to the lower bound question,
we consider the function
\begin{equation} \label{func:f*}
f^*= \bigvee_{i=1}^{s} (x_i \wedge x_0)
\vee \bigvee_{i=s+1}^{2s}
(x_i \wedge \overline{x}_0).
\end{equation}

Note that the set of minterms of $f^*$ is  exactly the union of the following sets
\begin{itemize}
\item  $\{\{x_i, x_j\} \mid 1 \leq i \leq s, \, s+1 \leq j \leq 2s\}$
\item  $\{\{x_0, x_i\} \mid 1 \leq i \leq s\}$
\item  $\{\{\overline{x}_0, x_j\} \mid s \leq j \leq 2s\}$
\end{itemize}
and the largest maxtem is the set $\{x_1, \dots, x_{2s}\}.$
 Clearly $k(f^*) =2$, and $l(f^*) = 2s,$ whence
 $f^* \in {\cal Q}.$ Moreover, it holds that
$\gamma(f^*) \leq l(f^*)/2 + 1$.

In order to show this let us consider the algorithm {\sc Bf2} in Figure \ref{fig:algo}, where we let
 $V_1=\{x_1,\ldots,x_s\}$ and $V_2=\{x_{s+1},\ldots,x_{2s}\}$.
We have the following results that concludes our analysis.

\begin{figure}
\begin{center}
\small
\begin{tabular}{|p{14cm}|}
\hline

Algorithm {\sc Bf2}

\hspace{1cm} {\bf Let } $L$ be the list of variables sorted  in order of non-decreasing cost

\hspace{1cm} {\bf While} neither $x_0$ nor a variable that evaluates to 1 have been read

\hspace{2cm}   Read the next variable of $L$

\hspace{1cm} {\bf End While}

\hspace{1cm}   {\bf If } ($x_0$ is read  and $x_0(\sigma)=0$) or the 1-value variable
belongs to $V_1$

\hspace{2cm}  Remove from $L$ the variables in $V_1$

\hspace{1cm}  {\bf Else }

\hspace{2cm}  Remove from $L$ the variables of $V_2$

\hspace{1cm}  {\bf While } the value of $f$ has not been determined

\hspace{2cm}    Read the next variable of $L$

\hspace{1cm}  {\bf End While } \\
\hline

\end{tabular}
\normalsize
\end{center}
\caption{The algorithm in the proof of Proposition \ref{prop:C-upperbound}} \label{fig:algo}
\end{figure}

\begin{proposition} \label{prop:C-upperbound}
We have  $\gamma^{\mbox{\footnotesize\sc Bf2}}(f^*)\leq l(f^*)/2 + 1$.
\end{proposition}
\begin{proof}
Let us consider the case of an assignment $\sigma_1$ such that
$f(\sigma_1) = 1.$
Let $\{x, y\}$ be the
cheapest proof for $f^*(\sigma_1) =1,$ with $c(x) \leq c(y).$
Let $j$ be the number of variables read by the algorithm when
the condition of the first {\bf While } loop becomes false.
Then, the cost incurred by {\sc Bf2} is
\begin{equation} \label{eq:uno}
\frac{c^{f^*}_{\mbox{\footnotesize \sc Bf2}}(\sigma_1)}{c^{f^*}(\sigma_1)} \leq
\max_{1 \leq j \leq 2s} \frac{j c(x) + \min\{s+1, 2s+1 - j\} c(y)}{c(x) + c(y)} \leq
s+1 = l(f^*)/2 + 1.
\end{equation}

Let now $\sigma_0$ be an assignment such that $f^*(\sigma_0) = 0.$
Without loss of generality, let $\sigma_0$ be an assignment that
sets $x_0$ to $1.$ Therefore, because of  $f^*(\sigma_0) = 0$ it also follows that
$\sigma_0$ sets  to $0$ the whole set $V_1$.
It is not hard to see that the only possible proofs are either $V_1 \cup \{x_0\}$ or $V_1 \cup V_2.$
The worst case for the algorithm {\sc Bf2} happens when it reads the whole set of variables $V_2$ and
$x_0$ before finishing to read $V_1$.
Let $P$ be the cheapest proof, and  $x^*$ be the variable of $P$  with the largest cost.
Since the algorithm reads the variables in order of increasing cost, we can bound the cost spent by
{\sc Bf2} as $c(P \setminus \{x^*\}) + (s+1) \times c(x^*).$ Therefore we have
\begin{equation} \label{eq:zero}
\frac{c^f_{\mbox{\footnotesize \sc Bf2}}(\sigma_0)}{c^f(\sigma)} \leq
\frac{c(P\setminus\{x^*\}) + (s+1) \times c(x^*)}{c(P \setminus \{x^*\}) + c(x^*)}\leq s+1  \leq l(f^*)/2 + 1.
\end{equation}

Taking the maximum of the ratios in (\ref{eq:uno})-(\ref{eq:zero}) we have the
desired result.
\end{proof}

\section{The class of symmetric Boolean functions}

\begin{sloppypar}The results of the previous section showed that for arbitrary Boolean functions, we can have $\textit{PROOF}(f) \neq  \gamma(f).$ We will show in this section that the difference between these two quantities can be arbitrarily large. More specifically, by considering the class of symmetric Boolean functions we will see that $\textit{PROOF}(f)$ is not bounded from above by any function of $\gamma(f)$. The complexity of symmetric functions
has long been studied in different models of computation~\cite{MP,W1984,Bar,Wegener-sym}.
\end{sloppypar}

For the class of symmetric Boolean functions we are able to provide exact evaluation of both
the competitive ratio and the extremal competitive ratio.

A Boolean function $f:\B^n\to \B$ is {\em symmetric} if, for every input vector $x\in \B^n$ and
for every permutation $\pi$ of the set $\{1,\ldots, n\}$, it holds that
$f(x_1,\ldots, x_n) =
f(x_{\pi_1},x_{\pi_2},\ldots, x_{\pi_n})$.
Equivalently, the value of $f$ on an input vector $x$ is fully determined by the number of 1's in~$x$.

There is a bijective correspondence between the set of all symmetric Boolean functions $f$ over $\B^n$
and the set of all mappings $\hat{\bf f}:\{0,1,\ldots, n\}\to \B$.
The correspondence is given by the relations $f(x) = \hat{\bf f}(\sum_{i = 1}^nx_i)$ and $\hat{\bf f}(k) = f(x^{(k)})$
where $x^{(k)}\in \B^n$ is any 0/1-vector containing precisely $k$ 1's, for example $x^{(k)}_i = \left\{
                                               \begin{array}{ll}
                                                 1, & \hbox{if $i\le k$;} \\
                                                 0, & \hbox{otherwise.}
                                               \end{array}
                                             \right.$
In what follows, we will always write $\hat{\bf f}$ for the mapping defined as above,
corresponding to a given symmetric Boolean function $f$.

\begin{sloppypar}
In order to state the competitive ratio of symmetric functions, we need some definitions.
A {\em block} of a symmetric Boolean function $f:\B^n\to\B$ is a maximal interval
\hbox{$[\ell,u] := \{\ell,\ell+1,\ldots, u-1,u\}$} on which the value of $\hat{\bf f}$ is constant.
Formally, $[\ell,u]$ is a block if $\ell,u\in \{0,1,\ldots, n\}$, $\ell\le u$, such that the following properties hold:
\begin{itemize}
  \item $\hat{\bf f}(k) = \hat{\bf f}(\ell)$ for all $k\in \{\ell,\ell+1,\ldots, u\}$,
  \item either $\ell = 0$ or $\hat{\bf f}(\ell-1)\neq \hat{\bf f}(\ell)$,
  \item either $u = n$ or $\hat{\bf f}(u+1)\neq \hat{\bf f}(u)$.
\end{itemize}
The {\em width} of a block $[\ell, u]$ is defined as the number $u-\ell+1$ of elements in the block. The {\em spread} of $f$, denoted by $s(f)$, is defined as the maximum width of a block of $f$. Note that the blocks of $f$ form a  partition of the set $\{0,1,\ldots,n\}$.
\end{sloppypar}

In the following proposition, we state a necessary and sufficient condition for
a restriction of a symmetric function to be constant, which can be useful for determining whether a given set of variables constitutes a proof for a given assignment, and will be implicitly used
in the proof of Theorem~\ref{theorem:comp_ratio} below.

\begin{proposition}
Let $f$ be a symmetric Boolean function over $V = \{x_1,\ldots, x_n\}$, let $U\subset V$ and let $\sigma_U$ be an assignment of values to the variables in $U$. Let $n_0$ ($n_1$) denote the number of variables in $U$ that are assigned 0 (1).

Then, the value of  $f$ is determined by the partial assignment
$\sigma_{U}$ if and only if the block $[\ell,u]$ of $f$ containing $n_1$ satisfies $u \ge n-n_0$.
\end{proposition}

\begin{proof}
Let $\sigma:V\to \mathbb{B}$ be any assignment that agrees with $\sigma_U$ on $U$.
Suppose that the block $[\ell,u]$ of $f$ containing $n_1$
satisfies $u \ge n-n_0$.
Then $\sigma$ assigns the value of 1 to at least $n_1\ge \ell$ and to at most $n-n_0\le u$ variables, independently of the values of variables in $V\backslash U$. Consequently, $f(\sigma) = \hat{\bf f}(\ell)$.

Conversely, suppose that the value of  $f$ is determined by the partial assignment
$\sigma_{U}$, and also that the block $[\ell,u]$ of $f$ containing $n_1$
satisfies $u < n-n_0$. Let $\sigma_0$ denote the assignment obtained from $\sigma_U$ by assigning 0 to all the variables in $V\backslash U$, and $\sigma_1$ the assignment obtained from $\sigma_U$ by assigning 1 to precisely $u+1-n_1$ variables in $V\backslash U$
(and 0 to the remaining ones). Then, we see that $f(\sigma_0) = \hat{\bf f}(n_1)\neq \hat{\bf f}(u+1)=f(\sigma_1)$ since
$n_1$ and $u+1$ belong to two consecutive blocks.
\end{proof}

\remove{
\subsection{Extremal competitive ratio}

\begin{theorem}
Let $f:\B^n\to \B$ be a non-constant symmetric Boolean function. Then $\gamma(f) = s(f)$.
\end{theorem}

\begin{prf}
First, we will show that $\gamma(f)\ge s(f)$. Consider an algorithm $\mathbb{A}$ for evaluating $f$.
Let $[\ell,u]$ be a block of $f$ of maximum width such that $u$ is as small as possible. Then, $u-\ell+1 = s$.
Moreover, without loss of generality we may assume that $u\le n-1$. (If $u=n$, then
since $f$ is non-constant, it holds that $\ell\ge 1$, and arguments similar to the ones below
would establish the same lower bound for this case.)

Then, $u-\ell+1 = s(f)=: s$.  Partition the set $V$ of variables into three pairwise disjoint sets $V_1$, $V_2$ and $V_3$ such that $|V_1| = s$, $|V_2| = \ell$ and $|V_3| = n-u-1$. Set $c(x) = 1$ for all $x\in V_1$, and $c(x) = 0$ for all $x\in V\backslash V_1$.
Every variable in $V_2$ is set to 1, every variable in $V_3$ is set to 0,
and every variable in $V_1$ is set to 1 except for the last variable in $V_1$ read by $\A$, which is set to 0.

On the given assignment, the function takes value $\hat {\bf f}(u)$. The cheapest proof is of cost 1 (it consists of the set $V_2\cup V_3\cup \{x\}$ where $x$ is the unique variable from $V_1$ that is set to 0). However, the algorithm must read all the variables from $V_1$ since otherwise it could not distinguish between the case when $f$ takes the value $\hat {\bf f}(u)$ or $\hat {\bf f}(u+1)$. This shows that $\gamma(f)\ge s(f)$.

We will now show that this lower bound is achieved by any greedy algorithm $\G$ that sorts the variables according to nondecreasing costs $c_{1}\le c_{2}\le \cdots \le c_{n}$ and reads the variable one by one in the order $(x_{1},\ldots, x_{n})$ as long as  the value of the function is not determined. Consider an arbitrary assignment $\sigma:V\to\B$, and let $r$ be the number of variables with value 1. Let $[\ell,u]$ be the block of $f$ containing $r$, and let $k$ denote the number of variables read by $\G$ for the assignment $\sigma$.

For $j=1,\ldots, n$, let $d_r=\sum_{i = 1}^jc_{i}$ denote the sum of the $j$ cheapest costs.

Since every proof for the value of $f$ under $\sigma$ contains at least $n-u$ variables of value 0 and at least $\ell$ variables of value 1, we have $k\ge n-u+\ell\ge n-s+1$. The total cost paid by $\G$ is $c^f_{\G}(\sigma)=d_k$. By the definition of $k$, the value of $f$ is not determined by the values of the $k-1$ cheapest variables. Therefore the cheapest proof for the value of $f$ under $\sigma$ costs at least
$d_{n-u+\ell-1}+c_{k}\ge d_{n-s}+c_{k}$.

It follows that for every cost function and for every assignment $\sigma$ there is a $k>n-s$ such that $c^f_{\G}(\sigma) \leq (k-n+s)\cdot c^f(\sigma)\leq s \cdot c^f(\sigma)$.
This shows that $\gamma^{\G}(f)\le s$, and consequently $\gamma(f)\le s$.\qed
\end{prf}

\medskip
\noindent
{\bf Remark:} an algorithm with optimal competitiveness can also be achieved via
the Linear Programming Approach~\cite{CicLab_icalp}.
}

\subsection{Competitive ratio for a given cost function}

\begin{theorem} \label{theorem:comp_ratio}
Let  $f:\B^n\to \B$ be a non-constant symmetric Boolean function with spread~$s$.
The competitive ratio of $f$ with respect to a cost function $c:V\to \mathbb{R}^+$ is given by
$$\gamma_c^{f} = \max_{k\,>\,n-s} \left\{\frac{d_k}{d_{n-s}+c_{k}}\right\}\,,$$
where $c_{1}\le c_{2}\le \cdots \le c_{n}$ are the variable
costs sorted in a non-decreasing order, and $d_k=\sum_{i = 1}^kc_{i}$ denotes the sum of $k$ cheapest costs.
\end{theorem}

\begin{prf}
Consider an algorithm $\mathbb{A}$ for evaluating $f$ with respect to a given cost function $c:V\to \mathbb{R}^+$. First, we will show that $\gamma^\A_c(f)\ge \max_{k\,>\,n-s} \left\{\frac{d_k}{d_{n-s}+c_{k}}\right\}$, by describing an adversary strategy for constructing an assignment $\sigma_c^\mathbb{A}$ which is `bad' for $\mathbb{A}$. Let $[\ell,u]$ be a block of $f$ of maximum width such that $u$ is as small as possible. Then, $u-\ell+1 = s$.
Moreover, without loss of generality we may assume that $u\le n-1$. (If $u=n$, then
since $f$ is non-constant, it holds that $\ell\ge 1$, and arguments similar to the ones below
would establish the same lower bound for this case.)

\begin{sloppypar}
Let $k$ be an index where the maximum is attained in the above expression $\max_{k\,>\,n-s} \{\frac{d_k}{d_{n-s}+c_{k}}\}\,.$ Let $x_1, \ldots, x_n$ be the variables in non-decreasing order by cost.  The adversary responds 0 to queries about $x_1,\ldots,x_{n - u - 1}$, it responds 1 to the first $k - n + u$ queries about other variables, and 0 to the remaining queries. Consider the partial assignment when the algorithm has seen exactly $k - n + u$ variables set to 1. We extend this assignment by setting any unset variables of $x_1, \ldots, x_{n - u -1}$ to 0, setting the cheapest of the other unset variables to 0, and setting all other unset variables to 1.
\end{sloppypar}

Any proof must contain at least $n - u$ variables set to 0 and $\ell$ variables set to 1.  Therefore, the algorithm must eventually query $x_1, \ldots, x_{n - u - 1}$, at least $k - n + u$ variables for which the adversary responds 1 (without all these it will never see the last 0), and the other variable set to 0; since these are $k$ variables, the total cost is at least $d_k$.  However, the cheapest proof consists of $x_1, \ldots, x_{n - u - 1}$, the $\ell$ cheapest variables set to 1, and the cheapest other variable set to 0.  If the other variable set to 0 is one of $x_{n - u}, \ldots, x_{n - s}$, then the others of these variables are the cheapest $\ell$ variables set to 1 and, so, the cheapest proof has cost $d_{n - s}$.  Otherwise, $x_{n - u}, \ldots, x_{n - s - 1}$ are the cheapest $\ell$ variables set to 1; since the cheapest other variable set to 0 is the cheapest unset variable remaining when the algorithm has queried at most $k - 1$ variables, it has cost at most $c_k$; therefore, the cheapest proof has cost at most $d_{n - s - 1} + c_k$. Therefore, if $\A$ is $\rho$-competitive, then $\rho\ge \frac{d_k}{d_{n-s}+c_{k}}$.
This shows that $\gamma_c^{f} \ge \max_{k\,>\,n-s} \{\frac{d_k}{d_{n-s}+c_{k}}\}\,.$

We will now show that this lower bound is achieved by any greedy algorithm $\G$ that sorts the variables according to nondecreasing costs $c_{1}\le c_{2}\le \cdots \le c_{n}$ and reads the variables one by one in the order $(x_{1},\ldots, x_{n})$ until the value of the function is  determined. Consider an arbitrary assignment $\sigma:V\to\B$, and let $r$ be the number of variables with value 1. Let $[\ell,u]$ be the block of $f$ containing $r$, and let $k$ denote the number of variables read by $\G$ for the assignment $\sigma$.

Since every proof for the value of $f$ under $\sigma$ contains exactly $n-u$ variables of value 0 and exactly $\ell$ variables of value 1, we have $k\ge n-u+\ell\ge n-s+1$. The total cost paid by $\G$ is $c^f_{\G}(\sigma)=d_k$. By the definition of $k$, the value of $f$ is not determined by the values of the $k-1$ cheapest variables. Therefore the cheapest proof for the value of $f$ under $\sigma$ costs at least $d_{n-u+\ell-1}+c_{k}\ge d_{n-s}+c_{k}$.

\begin{sloppypar}
We have shown that for every assignment $\sigma$ there is a $k>n-s$ such that \hbox{$c^f_{\G}(\sigma) \leq \rho_k c^f(\sigma),$} where $\rho_k = \frac{d_k}{d_{n-s}+c_{k}}$.
This shows that $\G$ is $\rho$-competitive, where
\hbox{$\rho = \max\{\rho_k\,:\,k\,>\,n-s\}$}. Consequently, $\gamma_c^{f} \le \gamma_c^{\G}(f) \le \max_{k\,>\,n-s} \{\frac{d_k}{d_{n-s}+c_{k}}\}\,.$\qed
\end{sloppypar}
\end{prf}

\subsection{Extremal competitive ratio}\label{sect:extremal}

The above proof can also be used to provide an exact evaluation of the best extremal competitiveness, $\gamma(f),$  for a symmetric Boolean function
$f.$ We have the following.

\begin{cor}
Let $f:\B^n\to \B$ be a non-constant symmetric Boolean function. Then $\gamma(f) = s(f)$.
\end{cor}
\begin{proof}
With reference to the notation in the statement of Theorem \ref{theorem:comp_ratio}, it is enough to prove that (i) for any cost assignment
and any $k > n-s,$ we have $\frac{d_k}{d_{n-s}+c_{k}} \leq s$; (ii)  there exists a cost assignment
such that $\max_{k > n-s} \frac{d_k}{d_{n-s}+c_{k}} = s.$

For (i) we have
$$\frac{d_k}{d_{n-s}+c_{k}} = \frac{d_{n-s} + \sum_{j=n-s+1}^k c_{j}}{d_{n-s}+c_{k}}
\leq \frac{d_{n-s} + s c_k}{d_{n-s}+c_{k}} \leq s.$$

Moreover, by considering a cost assignment in which $n-s$ variables have cost $0$ and the remaining ones have cost $1,$ we
have (ii).
\end{proof}


\bigskip

\noindent {\bf Example}. If $f$ is the {\em parity function}, that is, $\hat{\bf f}(k) = k~(\textrm{mod}~2)$ for all $k$,
then $s(f)= 1$ and consequently the competitive ratio $\gamma_c^{f}$ is equal to 1 for every cost function $c$. In fact, every proof must contain all the variables, hence $\textit{PROOF}(f)=n$ while $\gamma(f) = 1$. This shows that, for general Boolean functions, $\textit{PROOF}(f)$ is not bounded from above by any function of $\gamma(f)$.

\section{Further directions}

In \cite{CicLab_icalp} the authors introduced a new approach for the design of competitive
algorithms for the function evaluation problem. This linear programming approach
(${\cal LPA}$)  depends on the choice of   feasible solutions for  the following linear program defined on the set of the proofs of the function to evaluate
$${\bf LP}_{\bf f}: \left\{~\textrm{Minimize }\sum_{x\in V}s(x):
 \sum_{x\in P}s(x)\ge 1 \textrm{ for every $P\in {\cal P}(f)$ and $s(x)\ge 0$, for
every }x\in V\right\},$$
where ${\cal P}(f)$ denotes the set of all proofs of $f.$

We shall now focus on the best possible implementation of the ${\cal LPA}.$ We shall call this algorithm
$\mathbb{LP}.$ The algorithm  $\mathbb{LP}$
consists of reading the variable
$x = {\rm argmin}_{v \in V} \frac{c(v)}{s(v)},$ where $s(v)$ is the value assigned to $v$ in an optimal solution of
${\bf LP}_{\bf f},$ and then recursing on the restriction, $f_x,$ of $f$ obtained by fixing the value read for $x.$ More generally, for a subset $Y \subset V$, let $f_Y$ denote the
restriction of $f$ obtained by fixing the values in $Y$.

Let  $\Delta(f) = \max_{Y \subset {V}} \left \{ \sum_{x \in
V \setminus Y} s^{*}_Y(x) \right \},$
where $s^{*}_Y(\cdot)$ denotes the optimal solution of
${\bf LP_{f_{Y}}}.$
By \cite[Lemma 1]{CicLab_icalp} one gets the following result on the competitiveness of
$\mathbb{LP}.$

\begin{lemma}\cite{CicLab_icalp}
For any function $f,$ it holds that $\gamma^{\mathbb{LP}}(f) \leq  \Delta(f). $
\label{lem:key}
\end{lemma}

In order to prove that for a given function $f$, the algorithm $\mathbb{LP}$ achieves  extremal competitiveness $K$,
it is then sufficient to prove that for any restriction $f'$ of the function there exists a  feasible solution to the the linear program $\bf LP_{f'}$ with objective value not exceeding  $K.$

Cicalese and Laber proved that for any  Boolean function $\Delta(f) \leq \textit{PROOF}(f)$ \cite{CicLab_icalp}, which, together with a lower bound from~\cite{CharikarEtAl02a}, implies that  $\gamma(f) = \Delta(f) = \textit{PROOF}(f)$ for any monotone Boolean function $f$. They also showed that for the function
\begin{equation} \label{func:g}
g= (z \vee x_1) \wedge (z \vee x_2) \wedge (\overline{z} \vee x_3) \wedge (\overline{z} \vee x_4)
\end{equation}
it holds that $\gamma(f) < \textit{PROOF}(f)$ and observed that
in fact for such function we have $\gamma(f) = \Delta(f) < \textit{PROOF}(f).$

In this section we shall  show that both the function $g$ of (\ref{func:g})  and the function $f^*$ of (\ref{func:f*}) belong to
a particular  class of non-monotone functions for which
we still have $\gamma(f) = \Delta(f).$ This provides some support to  the conjecture that this equality
holds for all Boolean functions.

We shall need the following easy fact.

\begin{proposition} \label{prop:nonred}
Let $f$ be a monotone Boolean function. Then, for every   minterm (maxterm) $C$ of $f$  and
for every  $x \in C$,
there is a maxterm (minterm) $C'$ of $f$ such that  $C \cap C' =\{x\}$.
\end{proposition}

\bigskip

\subsection{More about $\gamma, \, \Delta$ and $\textit{PROOF}$ for (non-monotone)
Boolean functions}

As we saw in Section 4, there are Boolean functions which are non-monotone and such that $\gamma(f) = \Delta(f) \ll \textit{PROOF}(f).$ This is also shown by the following sequence of examples whose analysis is possible via the more general Lemma \ref{lemma:Bool_gamma_1} below.

Fix positive integers $k$ and $t$ and let
$X = \{x_{i\,j} \mid i=1,\dots,t, \, j = 0,1,\dots, 2^{k}-1\}$ and $Z = \{z_{1}, \dots, z_{k}\}.$
For each $j=0,1,\dots, 2^{k}-1,$ and each $s=1,\dots, k$ let $\ell_{s}(j)$ be $z_{s}$ or $\overline{z_{s}}$
according as  the $s$th digit in the binary expansion of $j$ is 1 or 0.
For $i=1,\dots, t$ and $j=0, \dots, 2^{k}-1,$
let  $f_{i\,j} = x_{i\,j} \wedge \bigwedge_{s=1}^{k} \ell_{s}(j).$

Then, for the function  $f = \bigvee_{i=1}^{t}
\bigvee_{j=0}^{2^k-1} f_{i\, j},$
\remove{Create $t \times 2^k$ minterms.
Each minterm has size $k+1$ and consists of one of the $2^k$ combinations (positive form and negative form)
of the first $k$ variables and a variable from the set of positive variables.
In addition, ensure that every combination of the $k$ variables
appear in $t$ minterms.}
we have $k+t=\Delta(f)=\gamma(f) <<  \textit{PROOF}(f)=t \times 2^{k}$.

\begin{lemma} \label{lemma:Bool_gamma_1}
\begin{sloppypar}
Let $f$ be a Boolean function whose set of variables is given by $V = \{x_{1}, \dots, x_{t},
z_{1}, \dots, z_{k}\},$ and such that there is a DNF for $f$  where each
variable in $Z=\{z_{1},\dots, z_{k}\}$
appears both in negated and non-negated form, and each variable in
$X=\{x_{1}, \dots, x_{t}\}$ appears either only in non-negated form
or only in negated form.
For each ${\bf a}=(a_{1}, \dots, a_{k}) \in \{0,1\}^{k},$ let $f_{\bf a}$ be the restriction of
$f$ obtained by fixing $z_{i}=a_{i}, $ for each $i=1,\dots, k.$
Let $\Gamma(f) = \max_{{\bf a} \in \{0,1\}^{k}} \textit{PROOF}(f_{\bf a})$ and
$G=\{{\bf a} \in \{0,1\}^{k} \mid \textit{PROOF}(f_{\bf a}) = \Gamma(f)\}.$
\end{sloppypar}

If there exists an ${\bf a} \in G$ and  a minterm $C^{1}$
(respectively a maxterm $C^{0}$) for $f_{\bf a}$
such that $|C^{1}| = \Gamma(f)$ (resp. $|C^{0}| = \Gamma(f)$)
and for each ${\bf b} \in \{0,1\}^{k} \setminus \{{\bf a}\}$ and for each  minterm
$D^{1}$ (resp. maxterm $D^{0}$) for $f_{\bf b}$ it holds that
$var(D^{1}) \not \subseteq var(C^{1})$ (resp. $var(D^{0}) \not \subseteq var(C^{0})$),
then $\gamma(f) = \Delta(f) = \Gamma(f) + k.$
\end{lemma}
\begin{proof}
We start by showing that $\gamma(f) \geq k+\Gamma(f).$
W.l.o.g., we can assume that all variables in $X$ only appear in non-negated form.

For each $\a \in \{0,1\}^{k},$ let  $\z_{\a} = \ell(z_{1}) \wedge \cdots \wedge \ell(z_{k}),$
where $\ell(z_{i}) = z_{i},$ if $a_{i}=1$ and $\ell(z_{i}) = \overline{z_{i}},$
if $a_{i}=0.$
We can factorize $f$ as follows:
$$ \bigvee_{\a \in \{0,1\}^{k}} \left(\z_{\a} \wedge f_{\a}\right) .$$

For each $\a \in \{0,1\}^{k},$ let ${\cal P}^{1}_{\a}, \, {\cal P}^{0}_{\a}$
be the set of minterms and maxterm for
$f_{\a}$ respectively. Note that $f_{\a}$ is  monotone. Therefore we shall
identify the maxterms and minterms of $f_{\a}$ with their sets of variables.

\noindent
{\em Claim.} Let $C$ be a proof for $f$ with respect to some assignment
for which $f$ takes value $1.$ If $C \cap Z = \emptyset$ then
$C = \bigcup_{\a \in \{0,1\}^{k}} C_{\a},$ where $C_{\a} \in {\cal P}_{\a}^{1}.$

Let $C \cap Z = \emptyset$ and assume (for the sake of the contradiction) that
there exists an $\a \in \{0,1\}^{k}$ such that, for each $C^{1}_{\a} \in {\cal P}^{1}_{\a}$
it holds that $C_{\a}^{1} \setminus C \neq \emptyset.$ Thus, there exists
$C^{0}_{\a} \subseteq \bigcup_{C^{1}_{\a} \in {\cal P}^{1}_{\a}}
(C^{1}_{\a} \setminus C)$ which is a maxterm for $f_{\a}.$
Clearly $C^{0}_{\a} \cap C = \emptyset.$
This, together with $C \cap Z = \emptyset$ implies that for any assignment $\sigma$
such that  $z_{i}(\sigma) = a_{i},$ for $i=1,\dots, k$ and $x(\sigma)=0$ for each $x \in C_{\a}^{0},$
we have  $f(\sigma) = 0.$
On the other hand,  for any assignment $\sigma$ such that for each $x \in C, \, x(\sigma) = 1,$
we have $f(\sigma)=1.$  Therefore, there is an assignment $\sigma$ that
forces $f$ to evaluate to $0$ and to $1.$
This absurdity proves that for each $\a,$ there exist $C_{\a} \in {\cal P}_{a}$,
such that $\bigcup_{\a \in \{0,1\}^{n}} C_{\a} \subseteq C.$
Moreover, by the minimality of $C,$ the inclusion cannot be proper.
The proof of the claim is complete.

Let ${\bf a} \in G$ and $C^{1}$ be a minterm\footnote{The proof for the case of a
maxterm is perfectly symmetric.}
for $f_{\bf a}$ such that $|C^{1}| = \Gamma(f)$ and
for each  ${\bf b} \in \{0,1\}^{k} \setminus \{{\bf a}\} $ and for each  minterm
$D^{1}$  for $f_{\bf b}$ it holds that
$D^{1} \not \subseteq C^{1}.$

Given a deterministic  algorithm $\cal A,$ we shall give a cost function and
an assignment $\sigma^{\cal A}$ which  forces $\cal A$ to incur the desired
competitive ratio.

To this end we set $c(z_{i}) = 1$ for each $i=1,\dots, k.$
For each $x \in C^{1}$ we set
$c(x)=1$ and for each $x \in X \setminus C^{1}$ we set $c(x) = 0.$

For each $x \in  X \setminus C^{1}$ we set $x(\sigma) = 0.$ Now let $i^{*}$ be such that
$z_{i^{*}}$ is the  last
variable in $Z$ probed by the algorithm $\cal A$ and let $x^{*}$ be the last variable in
$C^{1}$ read by the algorithm $\cal A.$
For each $x \in C^{1}\setminus \{x^{*}\}$ we set $x(\sigma) = 1.$
For each $i \in \{1,\dots, k\} \setminus \{i^{*}\}$ we set $z_{i}(\sigma)=a_{i}.$
We now consider two cases:

\medskip

\noindent
{\em Case 1.} {\cal A} probes $z_{i^{*}}$ before probing $x^{*}.$
Then, we set $z_{i^{*}}(\sigma) = a_{i^{*}}$ and $x^{*}(\sigma) = 0.$

Clearly, for this assignment we have $f(\sigma)=0.$ In fact we have $\z_{\b} \wedge f_{\b}(\sigma_{X}) = 0,$
for each $\b \neq \a.$ Moreover, we have
$\z_{\a} \wedge f_{\a}(\sigma_{X}) = f_{\a}(\sigma_{X})=0,$
since, by Proposition \ref{prop:nonred}, there exists a  maxterm, $C^{0}_{\a},$
for $f_{\a},$ such that $C^{0}_{\a} \cap C^{1} = \{x^{*}\},$ and
clearly $x(\sigma)=0,$ for each $x \in C^{0}_{\a}.$

It is also not hard to see that the algorithm $\cal A$ only finds out the value of $f$ after probing
$x^{*}.$ Thus $\cal A$ incurs a cost equal to $k+ |C^{1}|.$

On the other hand, for each $\b \neq \a,$ there exists a maxterm $C^{0}_{\b}$ for $f_{\b}$
such that $C^{1} \cap C^{0}_{\b} = \emptyset.$ For otherwise, $C^{1}$ would contain a minterm
for $f_{\b},$ against our hypothesis. Thus, the assignment $\sigma$ above sets all variables in $C^{0}_{\b}$
to $0.$ Therefore, there is a proof for $\sigma$ consisting of the variables in
$C^{0} = \bigcup_{\b \in \{0,1\}^{k}} C^{0}_{\b}.$ By noticing that $C^{0} \cap C^{1} = \{x^{*}\},$
we have the desired result for this case.

\noindent
{\em Case 2.} {\cal A} probes $x^{*}$ before probing $z_{i^{*}}.$
Then, we set $z_{i^{*}}(\sigma) = 1-a_{i^{*}}$ and $x^{*}(\sigma) = 1.$

Again, we have $f(\sigma)=0.$ To see this, let $\b \in \{0,1\}^{k}$
be defined by $b_{i} = a_{i}$ for each $i \neq i^{*}$
and $b_{i^{*}} = 1-a_{i^{*}}.$ Proceeding analogously to the previous case we can observe
 that there exists a maxterm for $f_{\b}$ that has empty intersection with $C^{1}.$ All variables in such maxterm
are given value $0$ by the assignment $\sigma.$ Thus we have
$f(\sigma) = \z_{\b} \wedge f_{\b}(\sigma_{X}) = f_{\b}(\sigma_{X}) = 0.$

The algorithm $\cal A$ spends again $k+|C^{1}|.$ In fact, until the variable $z_{i^{*}}$ is read, it is not possible to
discriminate between the case $f(\sigma)= \z_{\a} \wedge f_{\a}(\sigma_{X})= f_{\a}(\sigma_{X})=1$ and
$f(\sigma)= \z_{\b} \wedge f_{\b}(\sigma_{X}) = f_{\b}(\sigma_{X}) = 0$ respectively given
by the possibilities $z_{i^{*}} = 1$ and $z_{i^{*}} = 0.$

On the other hand, proceeding like in the previous case, we can see that there exists a proof for $\sigma$
of cost $1$ which is given by the set $\{z_{i^{*}}\} \cup \bigcup_{\a' \in \{0,1\}^{k}\setminus\{\b\}}
C^{0}_{\a'}.$

The proof of the lower bound is complete.

For the upper bound, consider the following easy construction for a
feasible solution for ${\bf LP_{f}}.$
Set $s(z) = 1,$ for each $z \in Z.$
Moreover, for each $\a \in \{0,1\}^k,$
let $s_{\a}$ be an optimal solution to ${\bf LP_{f_{\a}}}.$
Now, set $s(x) = 1/2^k \sum_{\a \in \{0,1\}^k} s_{\a}(x),$ for each $x\in X$.
Then we have
$$
\sum_{v \in V} s(v) = \sum_{z \in Z} s(z) + \sum_{x \in X} s(x)
 = k + 1/2^{k} \sum_{\a \in \{0,1\}^k} \sum_{x \in X} s_{\a}(x)$$
 $$\hspace{0.5cm}\leq
 k + 1/2^{k} \sum_{\a \in \{0,1\}^k} \textit{PROOF}(f_{\a}) \leq k+ \Gamma(f)$$

 For the feasibility, it is easy to see that for each proof $C$ for $f,$ such that
 $C \cap Z \neq \emptyset,$ we have $\sum_{v \in C} s(v) \geq
 \sum_{v \in C \cap Z} c(v) = |C \cap Z| \geq 1.$

Conversely, let $C$ be a proof that does not contain any variable in $Z.$
W.l.o.g., let us assume that $C$ is a proof with respect to some assignment for which
$f$ takes value $1.$
By the above claim, $C = \bigcup_{\a \in \{0,1\}^{k}} C_{\a},$
where $C_{\a}$ is a minterm for $f_{\a}.$
Therefore, by the definition of $s(\cdot),$  we have
$\sum_{x \in C} s(x) = \sum_{x \in \bigcup_{\a \in \{0,1\}^{k}} C_{\a}} s(x) =
\sum_{\a \in \{0,1\}^{k}} 1/2^{k} \sum_{x \in C_{\a}} s_{\a}(x) \geq
\sum_{\a \in \{0,1\}^{k}} 1/2^{k} = 1,$
where the  last inequality follows because $s_{\a}$ is a feasible solution for ${\bf LP_{f_{\a}}}$.
\end{proof}

\subsection{Final remarks and open questions}

\medskip
In Section~\ref{sect:quadr} we have provided some initial results about the extremal competitiveness of quadratic Boolean functions. It would be interesting to have a complete characterization of the quadratic case, and, more generally, to try to examine the extremal competitive ratio of Boolean functions of bounded degree (that is, those Boolean functions that admit a DNF only containing terms with at most $k$ variables, for some fixed  $k$).

Beyond the bounded degree case, a general and seemingly far-reaching goal of this research area is  to achieve a good understanding of the extremal competitiveness of general Boolean functions. In particular, it would be interesting to determine whether some other parameter of Boolean functions besides $\textit{PROOF}(f)$ is meaningfully related to the extremal competitiveness. When restricted to monotone Boolean functions, the extremal competitiveness $\gamma(f)$ coincides with $\textit{PROOF}(f)$. Is there a combinatorial parameter
that not only agrees with $\textit{PROOF}(f)$ for monotone Boolean functions, but also agrees with $\gamma(f)$ for all Boolean functions $f$?

Another related issue is whether the linear programming approach is always optimal for Boolean functions. It is known that in general, the ${\cal LPA}$ does not achieve the optimal results
(this is the case, for example, for the problems of searching or sorting). However, for all Boolean functions with known extremal competitiveness, either monotone or not---including the symmetric functions (cf.~Section~\ref{sect:extremal})---there exists an implementation of the ${\cal LPA}$ that achieves the optimal competitiveness.

Last but not least, we find it an interesting question to determine whether the extremal competitiveness is a measure of complexity of Boolean functions according to the axiomatization of such measures, as given by Wegener~\cite{Wegener}. Out of the three defining axioms, the most intriguing one to verify seems to be the one requiring that the measure should only attain positive integer values.
Since this question is also related to the question above about a combinatorial description of the extremal competitiveness, we state it explicitly: {\em Is the extremal competitive ratio integer, for every Boolean function $f$?} As a matter of fact, we are not aware of any function, not even non-Boolean, with a non-integral extremal competitive ratio.

%
%
%
%
%
%
%
%
%
\bigskip

\noindent
{\bf Acknowledgements.} We would like to thank Rudolf Ahlswede and Evangelos Kranakis for having
directed our attention to some of the problems addressed in this paper.

\remove{

\bibliographystyle{abbrv}
\small
\bibliography{eclipse}

\begin{thebibliography}{10}
\bibitem{Neu_Net}
M.~Anthony.
\newblock {\em Discrete mathematics of neural networks: selected topics}.
\newblock Society for Industrial and Applied Mathematics, Philadelphia, PA,
  USA, 2001.

\bibitem{Reliab}
M.~O. Ball and J.~S. Provan.
\newblock Disjoint products and efficient computation of reliability.
\newblock {\em Operations Research}, 36(5):703--715, 1988.

\bibitem{Bar} D.A. Barrington.
\newblock Bounded-width polynomial-size branching programs recognize exactly those languages
in $NC^1.$
\newblock in {\em Proc. of STOC 1986}, pp. 1--5, 1986.

\remove{
\bibitem{Berge}
C.~Berge.
\newblock {\em Hypergraphs}, volume~45 of {\em North-Holland Mathematical
  Library}.
\newblock Elsevier, 1995.
}

\bibitem{Games}
J.~M. Bilbao.
\newblock {\em Cooperative Games on Combinatorial Structures}.
\newblock Kluwer Academic Publisher, Boston, 2000.

\bibitem{Blum} A. Blum.
\newblock Relevant examples and relevant features: Thoughts from computational learning theory.
\newblock in {\em Proc. of the AAAI Symposium on Relevance}. 1994

\bibitem{Blum2} A. Blum and P. Langley.
\newblock Selection of relevant features and examples in machine learning.
\newblock {\em Artificial Intelligence}, 97:245--271, 1997.

\bibitem{Bor99}
E.~Boros and T.~\"{U}nl\"{u}yurt.
\newblock Diagnosing double regular systems.
\newblock {\em Annals of Mathematics and Artificial Intelligence},
  26(1-4):171--191, 1999.

\remove{
\bibitem{Canteau}  A. Canteaut and M.  Videau.
\newblock  Symmetric Boolean functions.
\newblock {\em IEEE Transactions on  Information Theory},
51 (8): 2791--2811, 2005.
}

\bibitem{CharikarEtAl02a}
M.~Charikar, R.~Fagin, V.~Guruswami, J.~M. Kleinberg, P.~Raghavan, and
  A.~Sahai.
\newblock Query strategies for priced information.
\newblock {\em Journal of Computer and System Sciences}, 64(4):785--819, 2002.

\bibitem{CicLab}
F.~Cicalese and E.~S. Laber.
\newblock A new strategy for querying priced information.
\newblock In {\em Proceedings of the 37th Annual ACM Symposium on Theory of
  Computing}, pages 674--683. ACM, 2005.

\remove{
\bibitem{ESA2005}
F.~Cicalese and E.~S. Laber.
\newblock An optimal algorithm for querying priced information: Monotone
  {B}oolean functions and game trees.
\newblock In {\em Proceedings of the 13th Annual European Symposium on
  Algorithms}, volume 3669 of {\em Lecture Notes in Computer Science}, pages
  664--676. Springer, 2005.
}

\bibitem{CicLab_soda}
F.~Cicalese and E.~S. Laber.
\newblock On the competitive ratio of evaluating priced functions.
\newblock In {\em Proceedings of the Seventeenth Annual {ACM}-{SIAM} Symposium
  on Discrete Algorithms ({SODA}-06)}, pages 944--953, 2006.

\bibitem{CicLab_icalp}
F.~Cicalese and E.~S. Laber.
\newblock Function evaluation via linear programming in the priced information
  model.
\newblock In {\em Proceedings of the 35th International Colloquium on Automata
  Languages and Programming}, volume 5125 of {\em Lecture Notes in Computer Science}, pages
  173--185. Springer, 2008.

\bibitem{Cic-Mil} F.~Cicalese and M.~Milani\v{c}.
\newblock Competitive evaluation of threshold functions in the priced information model.
\newblock  {\em Annals of Operations Research}, in press (doi:10.1007/s10479-009-0622-4).

\bibitem{Crama-Hammer}
Y.~Crama and P.~Hammer.
\newblock {\em Boolean Functions: Theory, Algorithms, and Applications}.
\newblock To be published by Cambridge University Press, New York. (in preparation).

\bibitem{Davidson}
S.~B. Davidson, H.~Garcia-Molina, and D.~Skeen.
\newblock Consistency in a partitioned network: a survey.
\newblock {\em ACM Computing Surveys (CSUR)}, 17(3):341--370, 1985.

\bibitem{HeiWig91}
R.~Heiman and A.~Wigderson.
\newblock Randomized vs. deterministic decision tree complexity for read-once
  Boolean functions.
\newblock {\em Computational Complexity}, 1:311--329, 1991.



\bibitem{JayKumSiv03}
T.~S. Jayram, R.~Kumar, and D.~Sivakumar.
\newblock Two applications of information complexity.
\newblock In {\em Proceedings of the 35th Annual ACM Symposium on Theory of
  Computing}, pages 673--682, 2003.

\bibitem{Kaplan2005}
H.~Kaplan, E.~Kushilevitz, and Y.~Mansour.
\newblock Learning with attribute costs.
\newblock In {\em Proceedings of the 37th Annual ACM Symposium on Theory of
  Computing}, pages 356--365. ACM, 2005.

\bibitem{Molina:85}
H.~Garcia-Molina and D.~Barbara.
\newblock How to assigne votes in a distributed system.
\newblock {\em Journal of the ACM}, 32(4):841--860, 1985.

\bibitem{MP} D.E. Muller and F. Preparata.
\newblock Bounds on complexities of networks for sorting  and switching.
\newblock {\em Journal of the ACM}, 22:195--201, 1975.

\bibitem{HeiNewWig93}
R.~Heiman, I.~Newman, and A.~Wigderson.
\newblock On read-once threshold formulae and their randomized decision tree
  complexity.
\newblock {\em Theoretical Computer Science}, 107(1):63--76, 1993.

\bibitem{henderson:88}
P.~B. Henderson and Y.~Zalcstein.
\newblock A graph-theoretic characterization of the $\text{PV}_{\text{chunk}}$
  class of synchronizing primitives.
\newblock {\em SIAM Journal on Computing}, 6(1):88--108, 1977.

\bibitem{Sw_Net}
S.~T. Hu.
\newblock {\em Mathematical theory of switching circuits and automata}.
\newblock University of California Press, Berkely, Los Angeles, 1968.



\bibitem{karchmer}
M.~Karchmer, N.~Linial, I.~Newman, M.~Saks, and A.~Wigderson.
\newblock Combinatorial characterization of read-once formulae.
\newblock {\em Discrete Mathematics}, 114:275--282, 1993.

\bibitem{FOCS::SaksW1986}
M.~Saks and A.~Wigderson.
\newblock Probabilistic {Boolean} decision trees and the complexity of
  evaluating game trees.
\newblock In {\em Proceedings of the 27th {IEEE} Symposium on Foundations of
  Computer Science}, pages 29--38. IEEE Computer Society, 1986.

\bibitem{Snir:1985:LBP}
M.~Snir.
\newblock Lower bounds on probabilistic linear decision trees.
\newblock {\em Theoretical Computer Science}, 38(1):69--82, 1985.

\bibitem{Wegener-sym} I. Wegener.
\newblock The complexity of Symmetric Boolean Functions.
\newblock in {\em Proc. of Computation Theory and Logic},
Lecture Notes in Computer Science, vol. 270,  pp. 433-442, 1987.

\bibitem{Wegener}
I. Wegener.
\newblock {\em The complexity of Boolean functions}.
\newblock B. G. Teubner, and John Wiley \& Sons, 1987.

\bibitem{W1984} I. Wegener.
\newblock Optimal decision trees and one-time-only branching
programs for symmetric Boolean functions.
\newblock {\em Information and Control}, 62:129--143, 1984.


\end{thebibliography}
}

\end{document}